\documentclass[prb,aps,twocolumn,amsmath,amssymb,showpacs,superscriptaddress]{revtex4}
\usepackage{graphicx}
\usepackage{psfrag}
\usepackage{color}
\usepackage{subfigure}
\usepackage{amsmath}
\definecolor{dred}{rgb}{0.7,0.0,0.0}
\allowdisplaybreaks 
\usepackage{array}

\begin{document}

%
% Title Page
%

\title{Orbital weight redistribution triggered by spin order in the pnictides}

\author{M. Daghofer} 
\email{M.Daghofer@ifw-dresden.de} 
\affiliation{IFW Dresden, P.O. Box 27 01 16, D-01171 Dresden, Germany}
\author{Q.-L. Luo}
\affiliation{Department of Physics and Astronomy, University of Tennessee,
Knoxville, TN 37996-1200 and
\\Materials Science and Technology Division, ORNL, Oak Ridge,
TN 37831-6032, USA}
\author{R. Yu}
\affiliation{Department of Physics \& Astronomy, Rice University, Houston, Texas 77005,  USA}
\author{D. X. Yao}
\affiliation{Department of Physics and Astronomy, University of Tennessee,
Knoxville, TN 37996-1200 and
\\Materials Science and Technology Division, ORNL, Oak Ridge,
TN 37831-6032, USA}
\affiliation{School of Physics and Engineering, Sun Yat-Sen University, Guangzhou
510275, China}
\author{A. Moreo}
\author{E. Dagotto}
\affiliation{Department of Physics and Astronomy, University of Tennessee,
Knoxville, TN 37996-1200 and
\\Materials Science and Technology Division, ORNL, Oak Ridge,
TN 37831-6032, USA}

\date{\today}
%\maketitle

\begin{abstract}
%Orbital order and its impact on 
The one-particle spectral function and its orbital composition are
investigated in a three-orbital model for the undoped parent compounds
of the iron-based superconductors. In the realistic parameter regime,
where results fit experimental data best, it is observed that the magnetization in the $xz$ and $yz$
orbitals are markedly different and the Fermi surface presents
mostly $xz$ character, as recently observed in photoemission
experiments  [T. Shimojima {\it et al.}, Phys. Rev. Lett. {\bf 104}, 057002 (2010)].
Since the ferro-orbital order in this regime is at most a few percent,
these results are mainly driven by the magnetic order.
An analogous analysis for a five-orbital model leads to similar conclusions.
\end{abstract}
 
 
\maketitle

%\section{Introduction}

\begin{figure}
\includegraphics[width = 0.42\textwidth,trim = 10 20 20 40, clip]
{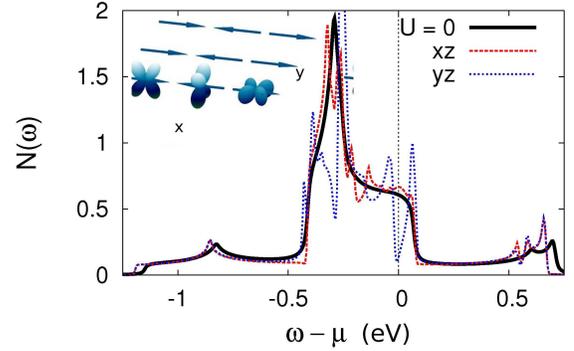}
\caption{(Color online) Density of states of the $xz$ and $yz$
  orbitals in the $(\pi,0)$-AF phase of the three-orbital model at
  $U=0.7$ and $J=U/4$. For these values of  
  $U$ and $J$, the orbital densities are $n_{xz}=n_{xz,\uparrow}+n_{xz,\downarrow}=1.590\approx n_{yz}=1.586$, 
and the magnetizations $m_{xz}=n_{xz,\uparrow}-n_{xz,\downarrow}=0.04\ll m_{yz}=0.15$. 
For $U=0$, $N(\omega)$ is identical for both  orbitals  (solid
line). A Gaussian broadening with $\sigma = 0.005$ was used. 
The inset illustrates the $(\pi,0)$-AF order considered here
  and the $xz$, $yz$, and $xy$ orbitals (left to right).\label{fig:dos_U35}}
\end{figure}

{\it Introduction.} In contrast to the cuprate superconductors, more than one band cross
the chemical potential in the pnictide
superconductors.\cite{LDA}%\cite{first,Singh:2008p1736,xu,cao,fang2, phonon0}
According to density functional theory (DFT) calculations, most of the
spectral weight at the Fermi surface (FS) arises 
from two of the iron $d$-orbitals, namely the $xz$ and $yz$ orbitals (inset of Fig.~\ref{fig:dos_U35}), which are
degenerate in the high-temperature tetragonal phase. Lowering the
temperature, the undoped parent compounds 
undergo a structural as well as a magnetic phase transition to an
orthorhombic phase with antiferromagnetic (AF)
order with wavevector $(\pi,0)$~\cite{sdw_n}%~\cite{sdw,neutrons1,neutrons2,neutrons3,neutrons4}
(inset of Fig.~\ref{fig:dos_U35}). This spin order breaks
the rotational symmetry of the original tetragonal lattice. In this regime, scanning
tunneling microscopy~\cite{Chuang:2010p2402} and resistivity measurements~\cite{rho_anisotropy} have indicated that
the electronic system presents symmetry-breaking properties
that far exceed the relatively modest difference of the
lattice constants before and after the transition.

To rationalize these results, orbital ordering has been suggested to occur together with the
magnetic ordering, lifting the degeneracy between the $xz$ and
$yz$ orbitals and inducing the lattice distortion.\cite{kruger:054504,ku,Lv10}
Such a ferro-orbital (FO) ordered state has 
anisotropic magnetic couplings, which stabilize the AF order
without necessarily frustrating any magnetic
interactions,\cite{ku,Daghofer_3orb,kruger:054504} thus
explaining the spin-wave dispersions which do not indicate frustrated 
AF couplings.~\cite{Zhao:2008PRL101p2397} 
However, while clear AF signatures have been observed in Angle Resolved 
Photoemission  (ARPES),~\cite{Lu:2008p2124,ARPES_el_pockets} the situation
is less clear for features due to orbital ordering. DFT results have
been interpreted in terms of FO order,\cite{ku} but a
detailed account of the impact of orbital order on the spectral
properties is still lacking. Finally, ARPES results can be fitted 
by DFT if the magnetic moment in the calculation  is 
artificially suppressed towards experimentally observed values, and
these calculations then yield a far smaller rearrangement of the two
hole pockets,~\cite{ARPES_el_pockets} indicating that FO order and
the ordered magnetic moment may be linked. 

In this paper, the spectral function of a three-orbital
model~\cite{Daghofer_3orb} is investigated with an emphasis on
features related to orbital polarization effects.%using the mean-field approximation.~\cite{Yu:2009p2127} 
This model allows for the stabilization of a regime with small ordered magnetic
moments~\cite{Yu:2009p2127,Daghofer:2008p1970} by selecting
intermediate values for the Hubbard repulsion.
% and calculate observables. 
One-particle spectral functions calculated in this regime
have already been shown to be qualitatively 
similar to experimental ones,~\cite{Daghofer_3orb} including the presence of
small hole- or electron-like extra pockets near the electron and hole
pockets of the uncorrelated bands.~\cite{ARPES_el_pockets,arpes3} 
We focus here on the orbital composition of the Fermi surface, which was not analyzed in those
previous investigations. It will be shown
that the FS has predominantly $xz$ character,  similar to recent experimental results
obtained with Laser-ARPES.~\cite{Shimojima:2010p2390} In this regime
with the polarized FS, the orbital magnetizations show a substantial difference between the
$yz$ and $xz$ orbitals, but there is hardly any static orbital
order. 
Our analysis of the three-orbital model is 
complemented by a discussion of a five-orbital
model,\cite{graser_5b} where similar results are found.
This model admits a regime with moderate FO order of $\approx 30\%$; the spectral
density and FS, however, more closely resemble ARPES results if the
FO is at most a few percent and the ordered magnetic moment is small or intermediate.~\cite{weak_AF}

%\section{Three- and Five- Orbital Models and Mean-field approach}\label{sec:model}

{\it Model.} The Hamiltonian studied here consists of the kinetic energy 
(tight binding) previously used for
three~\cite{Daghofer_3orb} or five~\cite{graser_5b} $d$ orbitals, as well
as the standard onsite Coulomb interaction terms comprised of the intraorbital
repulsion $U$, interorbital repulsion $U'$, and the $z$-component of the Hund's rule interaction
regulated by a coupling $J$, with $U=U'+2J$. The reader is referred to Refs.~\onlinecite{Daghofer_3orb,graser_5b} for more details. The overall electronic density per site 
$n$ is 4 (6) for the three (five) orbital model.
The spin-flip and pair-hopping terms,
%~\cite{Castellani:1978p1292} 
which are by symmetry
also part of the onsite interaction, drop out in our previous and current mean-field studies.~\cite{Oles_1983_Inter} The 
interacting Hamiltonian is then treated with a mean-field
approximation,~\cite{Yu:2009p2127} where we can compare a variety of phases with different
magnetic and orbital orders.~\cite{Daghofer_3orb} For a 
two-orbital model, our method was compared to the
Variational Cluster Approximation,~\cite{Aic03} and found to give
similar results.~\cite{Daghofer:2008p1970,Yu:2009p2127}
As previously reported, small to intermediate Coulomb repulsions $\lesssim 1 \textrm{eV}$
stabilize an AF metal in agreement with experiments.~\cite{Yu:2009p2127,Daghofer_3orb} 
Approximations beyond mean-field will likely increase the actual
values of $U$ and $J$ in the realistic regime.

{\it Results for three orbitals.} Figure~\ref{fig:dos_U35} shows the $xz$ and $yz$
contributions to the density of states, both for the AF metal found at  $U=0.7$ and
$J=U/4$, and for the uncorrelated system. In agreement with the
interpretation given to Laser-ARPES results in
Ref.~\onlinecite{Shimojima:2010p2390}, we find that most of the weight
at the FS arises from the $xz$ orbital in the AF state, while both
orbitals contribute equally in the tetragonal nonmagnetic state. The
total densities in the $xz$ and $yz$ orbitals are, however,
almost the same $n_{xz}=1.590 \approx n_{yz}=1.586$, 
i.e., there is (almost) no FO order. Only the states near the
chemical potential $\mu$ are $xz$-polarized, and a strong $yz$ peak at
energies $\approx -50\ \textrm{meV}$ approximately compensates 
for the missing $yz$-weight
around $\mu$. This peak comes from the opening of a gap that
stabilizes the AF order and that affects mainly the $yz$ portions of the
bands. The system remains metallic, because the $xz$ orbital does not
have a gap around $\mu$. The
stronger impact of the magnetic $(\pi,0)$-order on the $yz$ orbital also leads to 
a larger magnetization for this orbital, with $m_{yz}=0.15 \gg m_{xz}=0.04 \gg m_{xy}=0.014$. 
Such a larger value of $m_{yz}$ has been explained by a larger $yz$
hopping along the $x$-direction,~\cite{ku} but in the present
three-orbital model the $xz$ orbital has the largest hopping amplitude
along $x$.\cite{note_hopp_xy} While we have observed before a dominant 
$m_{xz}$ in the large-$U$ limit,~\cite{Daghofer_3orb} we attribute the
relatively large value of $m_{yz}>m_{xz}$ observed at smaller $U$ to the orbital character of the
electron pockets. The pocket found around $(\pi,0)$
in the uncorrelated model, with mostly $yz$ character, gets
folded into the central hole pockets and forms the gap mentioned
above, while the $xz$ pocket at $(0,\pi)$ is far less
affected.~\cite{note_orb_symm} Since the $yz$ orbital develops a
pseudogap at $\mu$ and the ungapped $xz$ orbital
consequently determines the states at the Fermi level, the slightly
higher resistivity in the ferromagnetic (FM) $y$-direction~\cite{rho_anisotropy} might
be due to the fact that the $xz$ orbital has larger hopping amplitudes
in the AF $x$-direction.

\begin{figure}
\subfigure{\includegraphics[width = 0.23\textwidth,trim = 110 10 100
  80, clip]{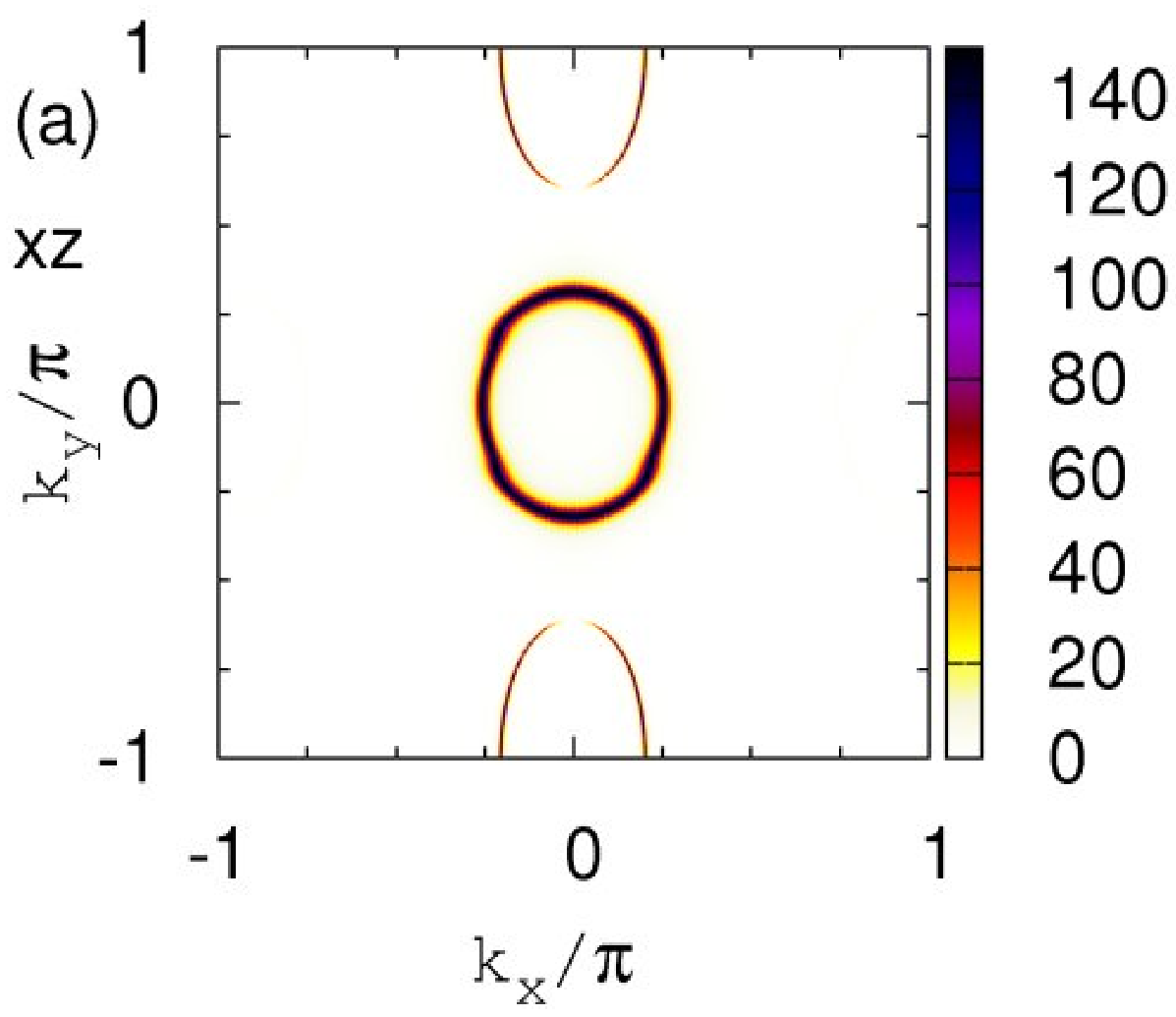}\label{fig:fs_3b_xz}}
\subfigure{\includegraphics[width = 0.23\textwidth,trim = 110 10 100
  80, clip]{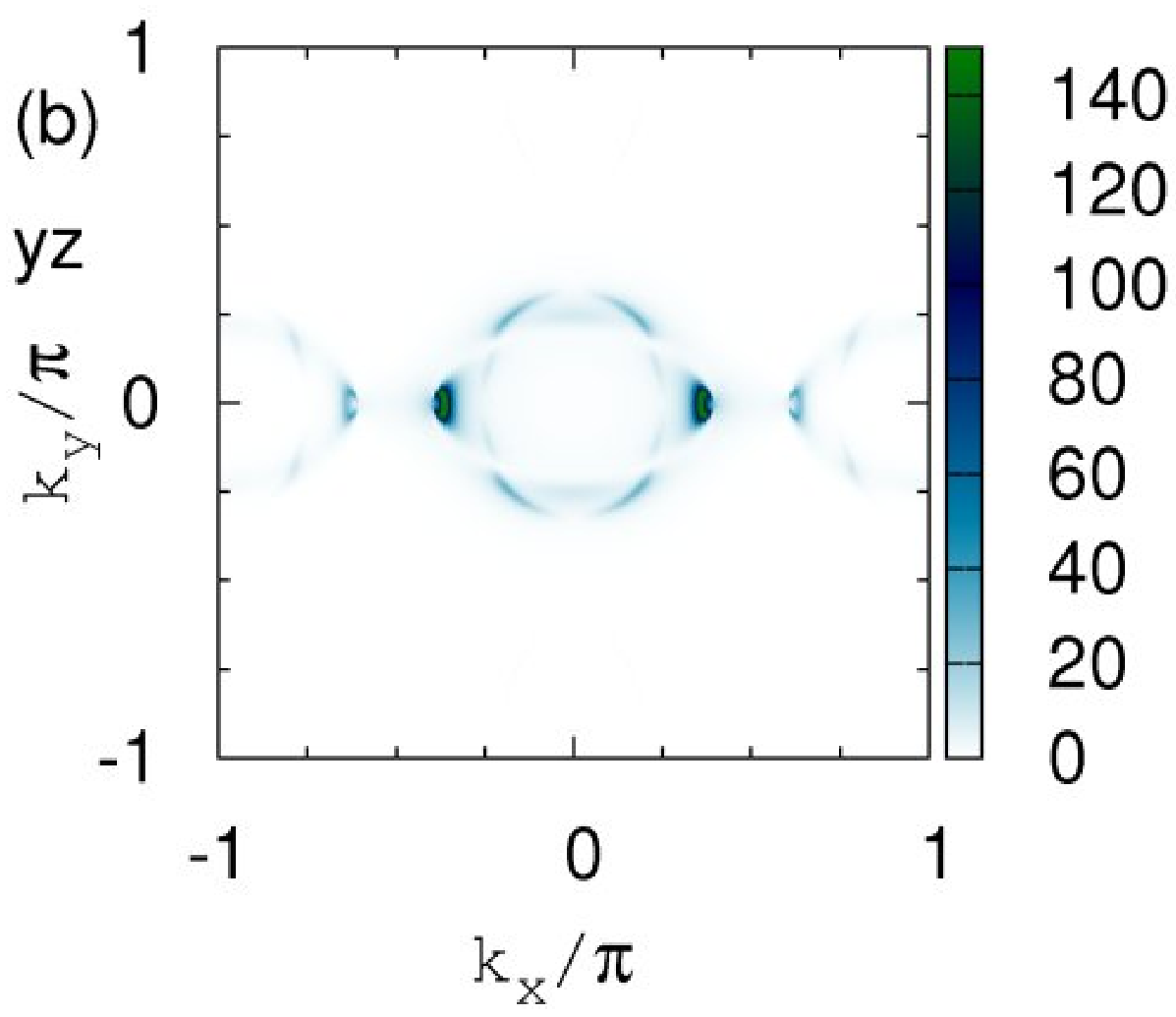}\label{fig:fs_3b_yz}}\\[-0.5em]
\subfigure{\includegraphics[width = 0.23\textwidth,trim = 110 10 100
  80, clip]{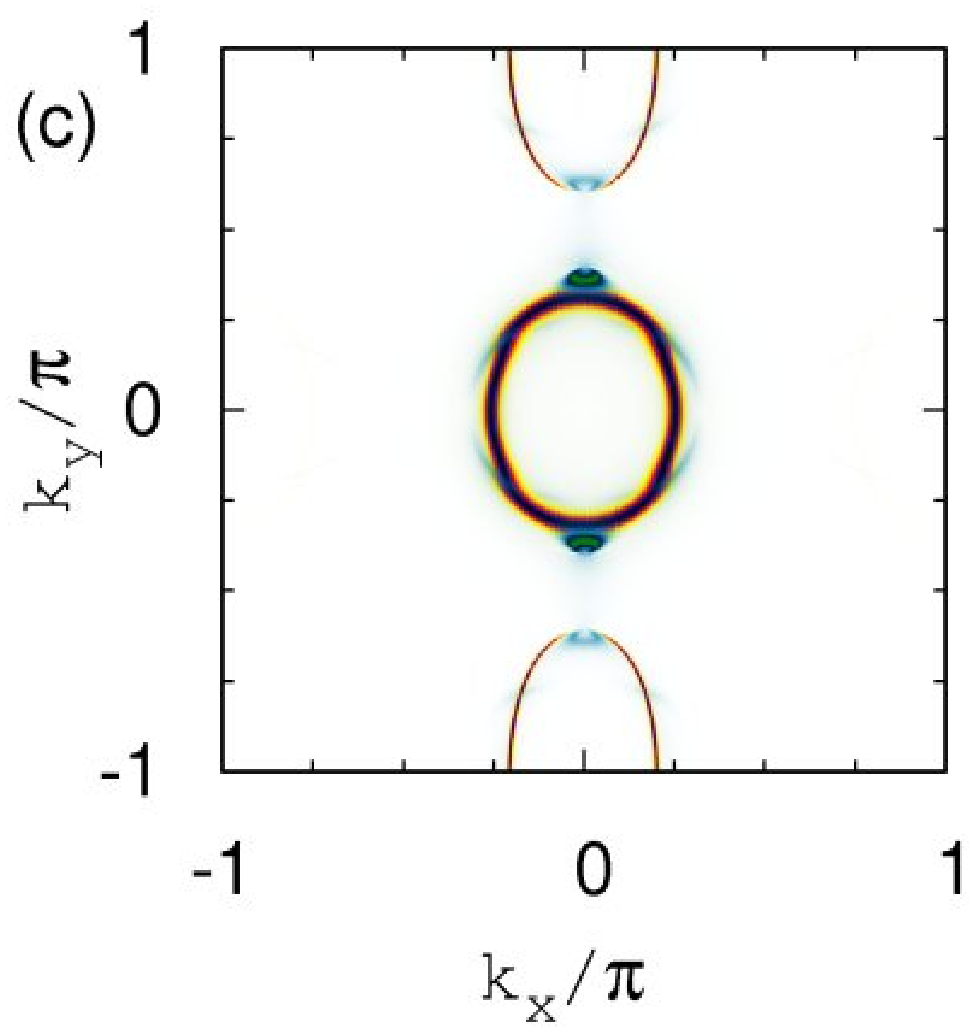}\label{fig:fs_3b_s}}
\subfigure{\includegraphics[width = 0.23\textwidth,trim = 110 10 100
  80, clip]{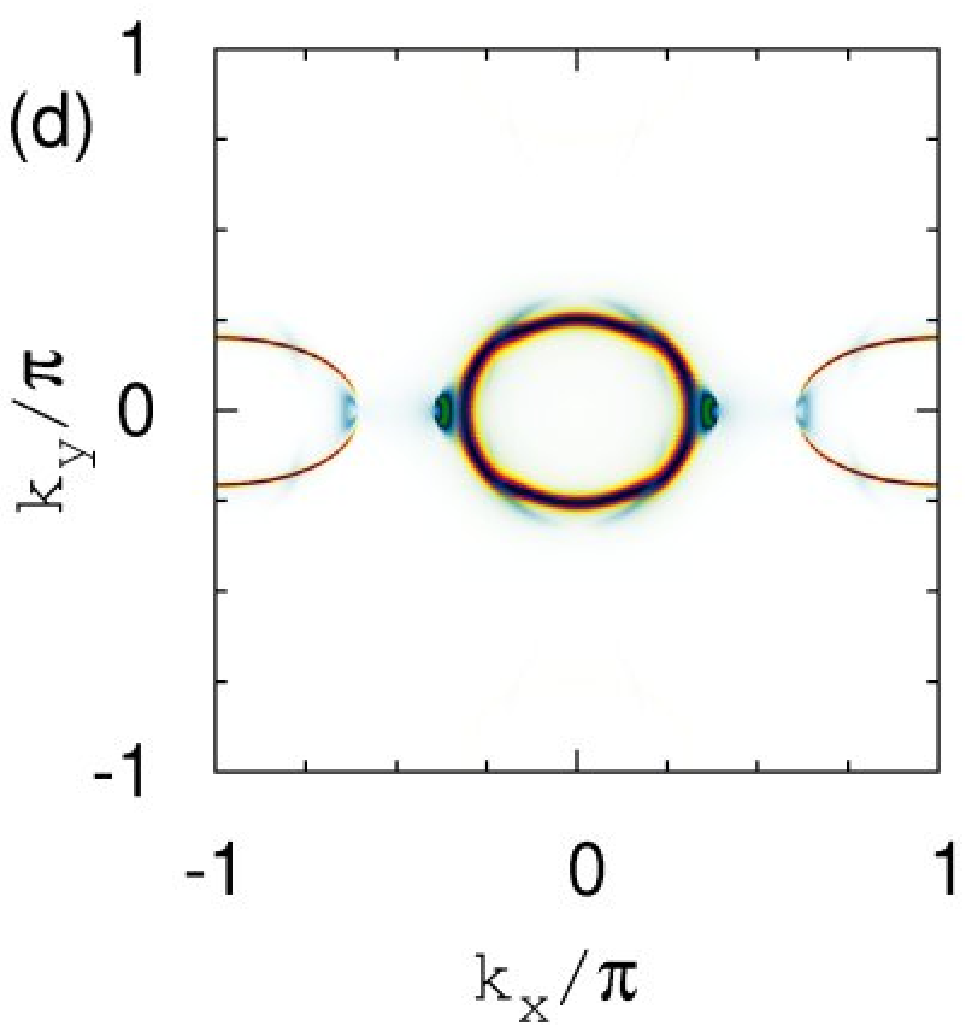}\label{fig:fs_3b_p}}\\[-0.5em]
\caption{(Color online) Contributions of the (a) $xz$ and (b) $yz$
  orbitals to
  the Fermi surface for the three-orbital model at the same parameters as in
  Fig.~\ref{fig:dos_U35}. (c) and (d) show the superpositions of $xz$
  and $yz$ as they would be expected to appear in the $s$ and $p$
  polarizations, see text. The FS also has a small weight in the $xy$
  orbital at the ``tip'' of the electron pocket (not shown).  
\label{fig:fs_U35}}
\end{figure}

Figures~\ref{fig:fs_U35}(a) and~(b) show the $xz$ and the $yz$
contributions to the Fermi surface for the same parameters as in
Fig.~\ref{fig:dos_U35}. As mentioned above, and as also reported
previously,~\cite{weak_AF} most of the FS is given by 
$xz$ states, but we also find small features coming from the $yz$
orbital. These small $yz$ electron-like pockets, see also Fig.~6(a) in
Ref.~\onlinecite{Daghofer_3orb}, are similar to 
$V$-shaped features reported in Laser-ARPES~\cite{Shimojima:2010p2390} and
their $yz$ character does not contradict the experimental findings: the laser spot is expected to catch
signals both from $(\pi,0)$- and $(0,\pi)$-ordered
domains, and the two polarizations pick up either the $xz$
or the $yz$ orbital. Since the $yz$ orbital takes the
same role for $(0,\pi)$ that $xz$ has for $(\pi,0)$, the
polarization sensitive to $xz$ symmetry is expected to find 
states with $xz$ character from $(\pi,0)$ domains together with features having
what corresponds effectively to `$yz$' from the rotated $(0,\pi)$
domains. Similarly, changing the polarization leads to $yz$ for $(\pi,0)$ plus `$xz$'
for $(0,\pi)$.~\cite{Shimojima:2010p2390} This situation can be
modeled by adding to the $xz$-weight at the FS the $yz$
contributions rotated by 90 degrees, because they stem
from domains with rotated AF order. Figures~\ref{fig:fs_3b_s}
and~\ref{fig:fs_3b_p} show the expected result for the two
polarizations, and one clearly observes the rotation of all features by
90 degrees, as seen in experiments. The
rotation would only break down for a hybridized FS that contains
substantial contributions from both the $xz$ and $yz$ orbitals. Both
hole pockets present such a mixture in the uncorrelated
bands,~\cite{Daghofer_3orb} but Figs.~\ref{fig:fs_U35}(a) and~(b)
clearly show that the AF order removes the hybridization and each
feature of the FS in the AF phase has (almost) only $xz$ \emph{or}
$yz$ character.

\begin{figure}
  \subfigure{\includegraphics[width=0.4\textwidth,trim = 0 60 10 30,clip]{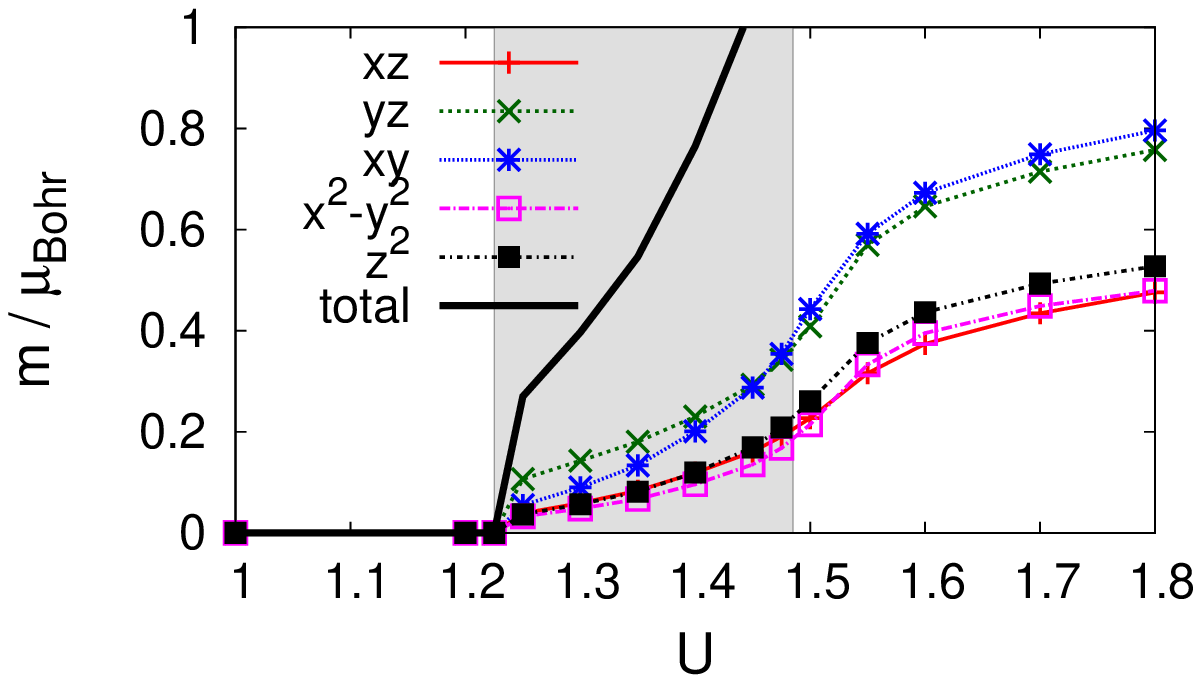}\label{fig:U_m}}\\[-1em]
  \subfigure{\includegraphics[width=0.4\textwidth,trim = 0 20 10 30,clip]{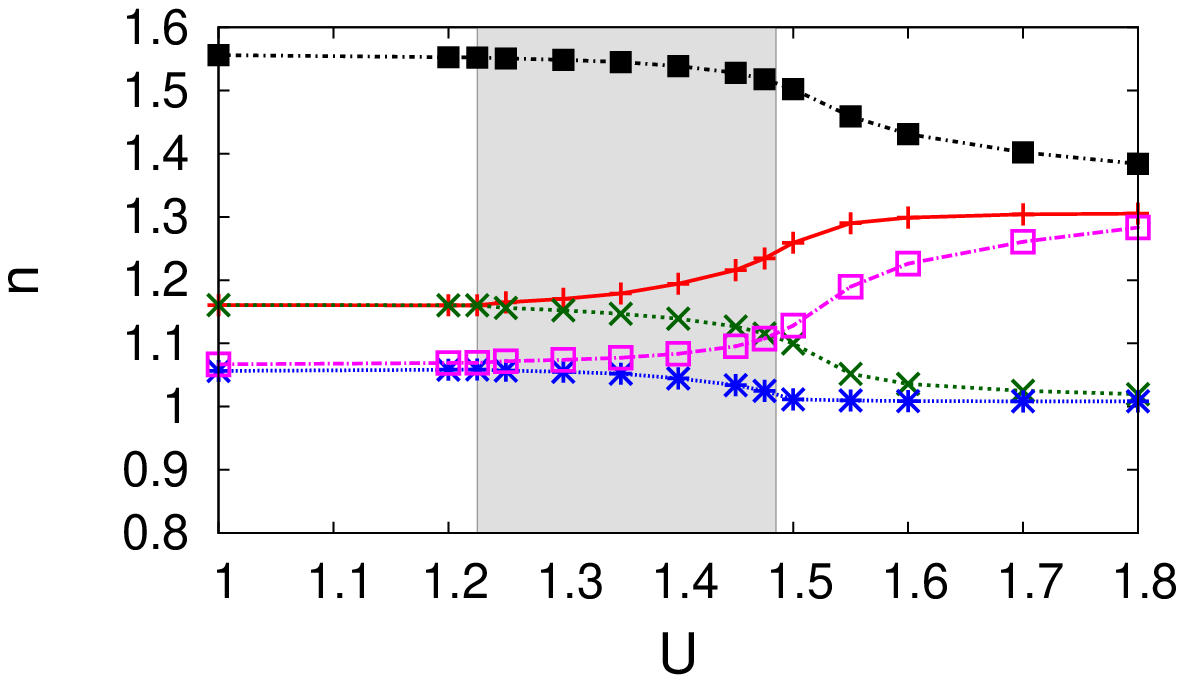}\label{fig:U_n}}
  \caption{(Color online) Orbital dependent (a) magnetizations (in
    $\mu_{\textrm{B}}$) and (b) electronic densities
    of the five-orbital model varying the strength
    of the Coulomb repulsion $U$, at $J=U/4$. In (a), the total magnetization is
    also given. The gray area denotes the regime with small ordered
    magnetic moment and almost no FO order. Its lower
    boundary is given by the onset of a finite ordered moment. Since
    orbital order develops more gradually, its upper boundary was
    defined as the inflection point of the orbital densities, which
    coincides well with an inflection point in the
    magnetization.\label{fig:U_m_n}} 
\end{figure}

{\it Results for five orbitals.} 
A similar analysis was carried out for a five-orbital
model\cite{graser_5b} and Fig.~\ref{fig:U_m_n} shows the mean-field
results for the 
magnetizations and densities in the five $d$-orbitals, varying the strength of the onsite Coulomb
repulsion $U$.\cite{comment}
A realistic
constant ratio $J=U/4$ was chosen, and it was checked that a slightly 
larger or smaller $J$ does not qualitatively alter our conclusions. For
small $U$, the system remains an uncorrelated metal without any
magnetic ordering and practically unchanged orbital densities. As was
reported for the other multi-orbital models,~\cite{Daghofer_3orb,Yu:2009p2127} AF order starts
to develop at a critical value of $U$, see Fig.~\ref{fig:U_m}. The staggered
magnetization per site, which corresponds to the ordered magnetic
moment, grows continuously in this regime and remains smaller than
$1.5 \mu_{\textrm{B}}$. 

Similarly as for the three-orbital model, the five-orbital model is away from
half-filling, and orbital ordering effects could therefore occur 
more easily than in the half-filled 
four- and two-orbital models.~\cite{moreo,kubo}
However, again similarly as for the three-orbital model, the first critical
$U$ turns out not to affect the orbital densities as strongly as the
magnetization, see Fig.~\ref{fig:U_n}. The
densities only slowly begin to vary after a robust magnetization has
set in and for a finite 
window in $U$ the difference remains in the low percent range, 
far smaller than the magnetization. Moreover, the difference
in orbital densities is also smaller than the difference in orbital
magnetizations with $m_{yz}> m_{xz}$, 
due to the $yz$ band being more strongly gapped
around the chemical potential (see the discussion for the
three-orbital model above). Only for larger values of $U$, where the
ordered magnetic moment is already quite large, a moderate
FO order sets in with $\approx 30\%$ more electrons in the $xz$
orbital. Even in this phase, we find that all orbitals are
affected to a similar degree, even though the $xz$ and $yz$ orbitals,
which are degenerate in the uncorrelated case and make up most of the
weight at the FS, might {\it a priori} be expected to be particularly
susceptible to symmetry-lowering orbital order.

\begin{figure}
  \subfigure{\includegraphics[width=0.42\textwidth,trim = 0 50 0 70,clip]{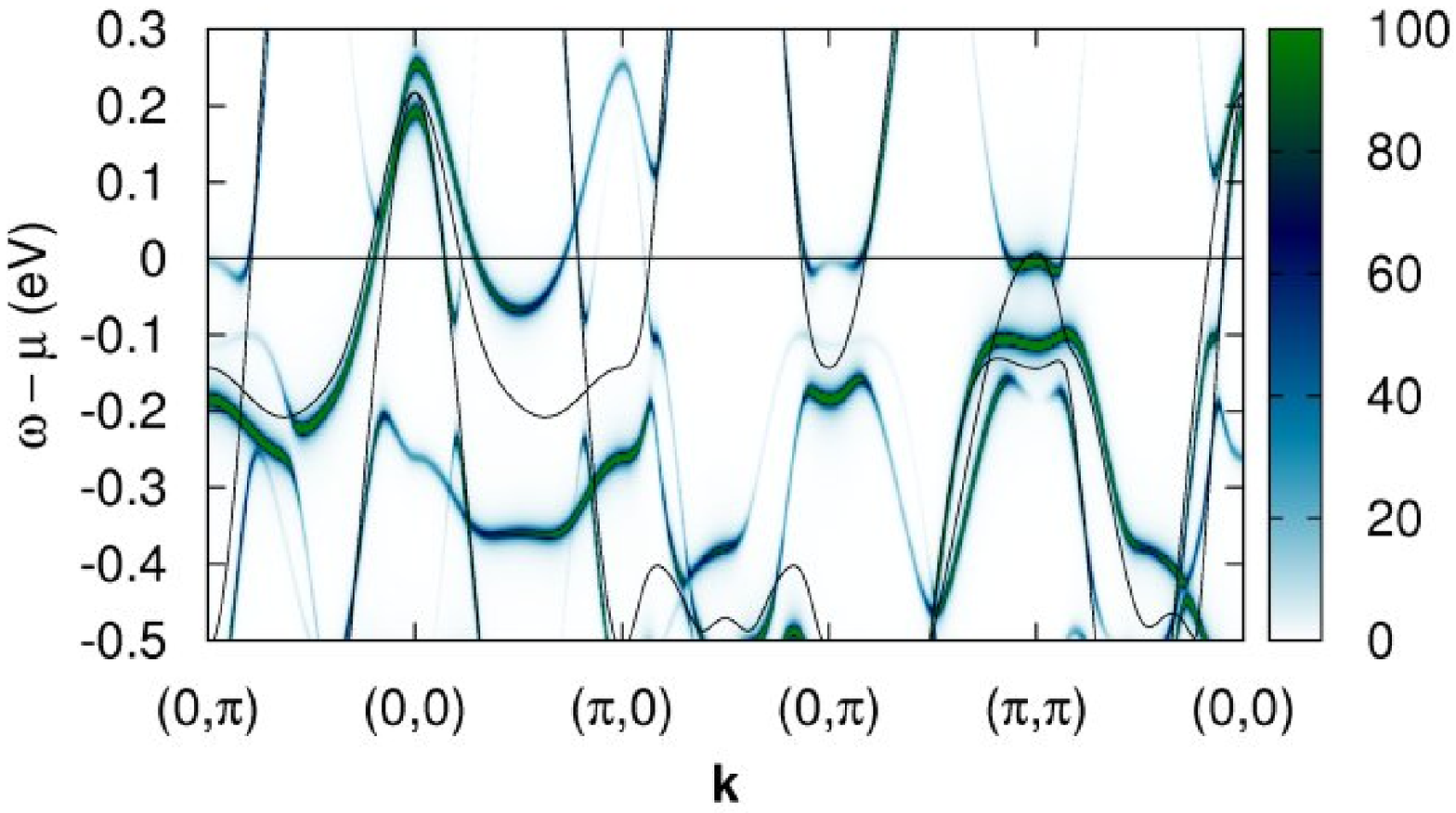}\label{fig:Ak_5b_U135}}\\[-0.5em]
  \subfigure{\includegraphics[width=0.42\textwidth,trim = 0 20 0 70,clip]{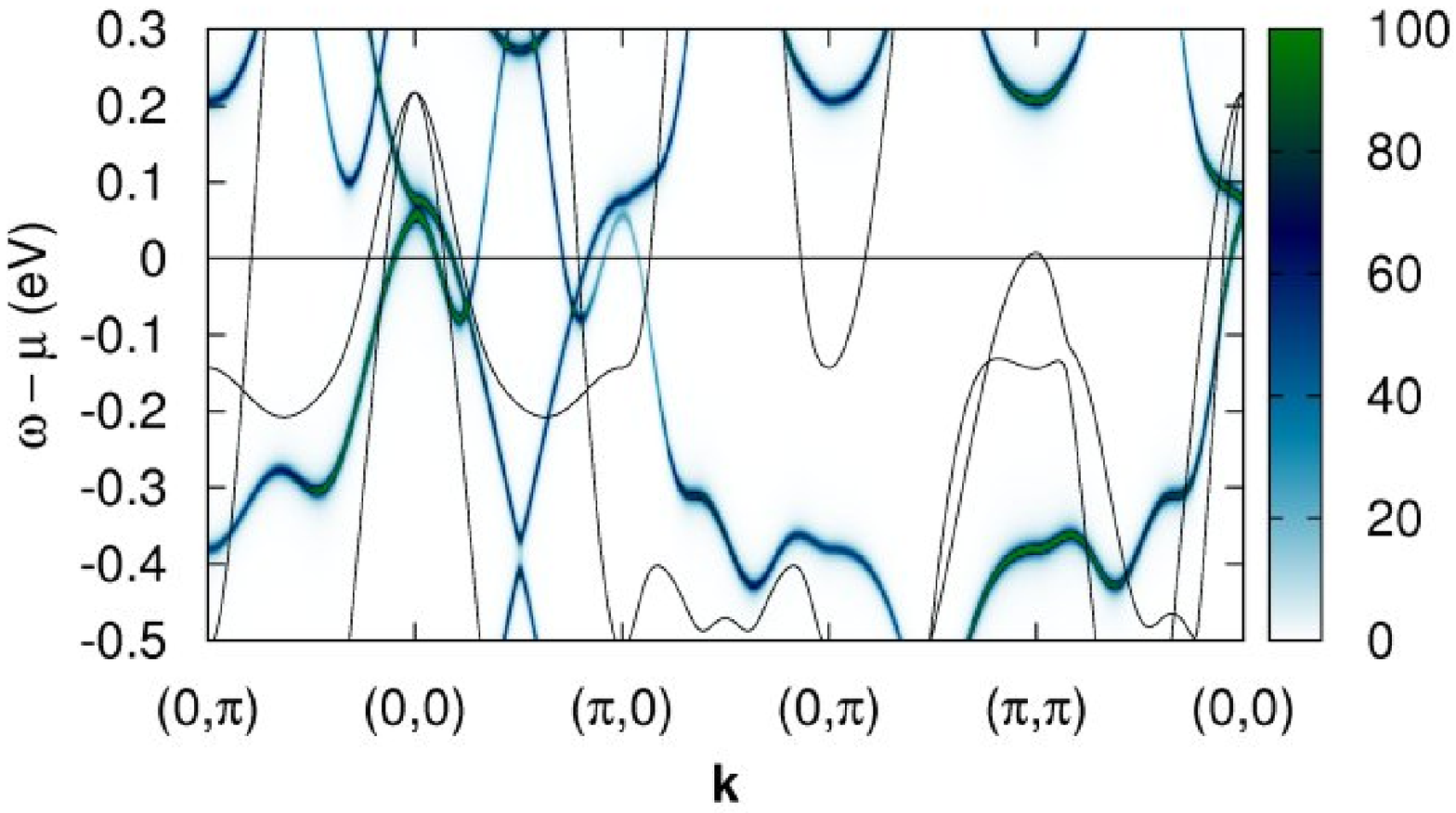}\label{fig:Ak_5b_U16}}
  \caption{(Color online) One-particle spectral function $A({\bf k},\omega)$ for
    (a) the AF metal without orbital order and (b) the AF
    metal with moderate FO order at $U=1.6$. 
    In (a), for $U=1.35$ and $J=U/4$, the total staggered magnetization is
    $m_{\textrm{tot}}=0.55 \mu_{\textrm{B}}$, and the orbital
    contributions of the $xz$ and $yz$ orbitals are $m_{xz}=0.08 \mu_{\textrm{B}}$ and
    $m_{yz}=0.18 \mu_{\textrm{B}}$. The difference in the densities
    is $n_{xz}-n_{yz}=0.032$. In (b), for $U=1.6$ and $J=0.4$,
    $m_{\textrm{tot}}=2.52 \mu_{\textrm{B}}$,
    $m_{xz}=0.37\mu_{\textrm{B}}$, and 
    $m_{yz}=0.65 \mu_{\textrm{B}}$; the difference in the densities
    is $n_{xz}-n_{yz}=0.26$. The thin black lines give the
    uncorrelated bands at $U=0$.\label{fig:Ak_5b} }
\end{figure}

Figure~\ref{fig:Ak_5b_U135} gives the spectral function $A({\bf k},\omega)$
for $U=1.35$ and $J=U/4$, where the AF ordered magnetic moment is
relatively small $m_{\textrm{tot}}= 0.55 \mu_{\textrm{B}}$ and where
the densities in the $xz$ and $yz$ orbitals differ only by a few
percent i.e., in the gray-shaded area of Fig.~\ref{fig:U_m_n}. The
spectral function is rather similar to the uncorrelated one, but some
gaps have opened, at the chemical potential (mostly for the $yz$ orbital) and
also away from it. The resulting ``shadow'' bands of magnetic origin~\cite{haas} 
form additional hole-pocket--like features next to the original electron pockets, in
agreement with ARPES.~\cite{arpes3} Similar to the three-orbital model
discussed above, $xz$-$yz$-hybridization of the hole pockets has given way to a
large $xz$-polarized central pocket and small satellites with $yz$
character. 
If a slightly larger $U=1.6$ is
chosen, so that the system develops some FO order, see
Fig.~\ref{fig:U_n}, the spectral function changes considerably, as it can be
seen in Fig.~\ref{fig:Ak_5b_U16}: all remnants of the
original electron pockets have completely disappeared, there are no longer
bands just below the chemical potential around $(\pi,0)$/$(0,\pi)$ as
seen in ARPES,~\cite{ARPES_el_pockets} and the features around $(0,0)$ and
$(\pi,0)$ are far more symmetric to each other than in ARPES.~\cite{ARPES_el_pockets,arpes3} 
The strong reconstruction into
features that resemble neither the uncorrelated bands nor ARPES
experiments arises because the interaction
is now strong enough to involve the states at $(\pi,\pi)$, located
just below the chemical potential at $U=0$, in the magnetic
order.

%\section{Conclusions}\label{sec:conclusions}

{\it Conclusions.} In summary, both the three- and five-orbital models have instabilities towards
orbitally ordered states, and the instability can be driven by
$(\pi,0)$-AF order. However, significant orbital order 
%only develops in the presence of 
requires 
a relatively strong onsite Hubbard repulsion $U$
and goes together with a large ordered magnetic moment and a significant
reconstruction of the one-particle bands and FS, see
Ref.~\onlinecite{Daghofer_3orb} and Figs.~\ref{fig:U_m_n} and~\ref{fig:Ak_5b_U16}. At
small to intermediate $U$, a realistic AF metal with small ordered
magnetic moments  is found where the densities in the $xz$ and $yz$
orbitals differ by at most a few percent. 

The orbital \emph{magnetization}, on the other hand, is far stronger for
the $yz$ orbital [for AF order with ordering vector $(\pi,0)$] than it
is for $xz$, which suggests that the orbital degree of freedom strongly
couples to the magnetic order. Such a more dynamic picture of the
orbital degree of freedom in pnictides is also corroborated by the
one-particle density of states, where the states near the Fermi
surface have more $xz$ character than $yz$, leading to a FS
with substantial orbital polarization, even in a regime where FO order
is at most a few percent. Another effect of the magnetic
order is the breakdown of the hybridization between the $xz$ and $yz$
orbitals: while the uncorrelated FS shows features with
mixed $xz$-$yz$ character, all features in the correlated FS are either
purely $xz$ or, for some smaller pockets, purely $yz$, in agreement
with Laser-ARPES results.~\cite{Shimojima:2010p2390}

%\begin{acknowledgments}
{\it Acknowledgments.} 
This research was sponsored by the NSF grant DMR-0706020, the
Division of Materials Science and Engineering, Office of Basic Energy Sciences,
U.S. DOE (A.M. and E.D.), and by the Deutsche
Forschungsgemeinschaft (DFG) under the Emmy-Noether program. We
acknowledge valuable discussions with H. Rosner, P. M. R. Brydon, and K. Koepernik.
%\end{acknowledgments}

% If you want to use bibtex, uncomment the following three lines and
% replace the \cite{LDA} and \cite{sdw_n} in the first paragraph by
% the full lists:
%\bibliographystyle{prsty-etal}
%\bibliography{feas_theory,feas_exp,notes,orb,vca}

\begin{thebibliography}{10}
\bibitem{note_orb_symm}
Since our mean-field approach breaks the orbital symmetry by neglecting the
  `pair-hopping' and spin-flip terms of the onsite interaction, we have to
  check that the $yz$ electron pocket's predominant interaction with the $yz$
  parts of the hole pockets is not an artifact of the approximation. We perform
  the mean-field calculation in a rotated basis with orbitals $|a\rangle=\cos
  \alpha |xz\rangle + \sin \alpha |yz\rangle$ and $|b\rangle= -\sin \alpha
  |xz\rangle + \cos \alpha |yz\rangle$~\cite{Daghofer_3orb} and find that (i)
  the energy is minimal for the original $\alpha=0$ and that (ii)
  $m_{yz}>m_{xz}$ persists for $\alpha \neq 0$.
\bibitem{comment}
For the intermediate-$U$ regime with small ordered moment, the $(\pi,0)$ phase
  has actually slightly higher energy than a $(\pi,\pi)$-AF phase with large
  ordered moment, but since the five-band model is widely used by several
  authors, we are nevertheless going to discuss its properties under the
  assumption of $(\pi,0)$ order.
\end{thebibliography}
%\end{document}

\end{document}